\documentclass[sigconf,authorversion,nonacm,screen]{acmart}

\AtBeginDocument{%
  \providecommand\BibTeX{{%
    \normalfont B\kern-0.5em{\scshape i\kern-0.25em b}\kern-0.8em\TeX}}}

\usepackage{soul}
\usepackage{balance}
\begin{document}

\title[Meaningful Play and Malicious Delight]{Meaningful Play and Malicious Delight:\\Exploring Maldaimonic Game UX}

\author{Katie Seaborn}
\email{seaborn.k.aa@m.titech.ac.jp}
\orcid{0000-0002-7812-9096}
\affiliation{
  \institution{Tokyo Institute of Technology}
  \city{Tokyo}
  \country{Japan}
}

\author{Satoru Iseya}
\orcid{0000-0002-0228-4671}
\email{iseya.s.aa@m.titech.ac.jp}
\affiliation{
  \institution{Tokyo Institute of Technology}
  \city{Tokyo}
  \country{Japan}
}




\renewcommand{\shortauthors}{Seaborn and Iseya}

\begin{abstract}
  Maldaimonia is a new experiential concept that refers to self-actuali\-zation and self-expression through egocentric, destructive, and/or exploitative activities. Still, it is unclear whether maldaimonia is an actual facet of real experience. As a subversive orientation, it may be rare or socially challenging to discuss openly. However, video games provide a space in which people can be expressive in different ways without the same repercussions as in real life. Indeed, game spaces may be one of the few contexts in which to study maldaimonic experiences. In this study, we examined whether and how maldaimonia exists as a feature of game user experiences by analyzing critical self-reports of gaming activities, confirming its existence. We contribute this new construct to work on "dark play" in games research.
\end{abstract}

\ccsdesc[500]{Human-centered computing~Human computer interaction (HCI)}
\ccsdesc[500]{Human-centered computing~Empirical studies in HCI}
\ccsdesc[500]{Applied computing~Computer games}

\keywords{Maldaimonia, game user experience, maldaimonic game UX, user experience, games}


\received{20 February 2007}
\received[revised]{12 March 2009}
\received[accepted]{5 June 2009}

\maketitle

\section{Introduction}

Human-computer interaction (HCI) has long recognized that we have a range of experiences with interactive devices, spaces, and agents, including games—some predictable, others not—that go beyond mere usability. Indeed, user experience (UX) has become central to the study and design of interactive experiences \cite{anderson_seductive_2011,hassenzahl_user_2008,hassenzahl_user_2006,korhonen_understanding_2009}. UX is a broad and imprecise \cite{forlizzi_understanding_2004} subjective factor that covers emotion and affect, experiential features, such as time and complexity, and non-instrumental impressions, including aesthetics and pleasure \cite{hassenzahl_user_2006}. Subsequent efforts have aimed for conceptual congruity and how to achieve “good” UX across a range of experiences that may not be simply “positive” \cite{hassenzahl_user_2008,hassenzahl_user_2006,korhonen_understanding_2009,wright_making_2004}.

One arm of this work has drawn from ancient yet persisting Greek philosophies of living the good life: \emph{hedonia} and \emph{eudaimonia} \cite{huta_pursuing_2013,deci_hedonia_2008}. Hedonia refers to an orientation towards pleasure-seeking and pleasurable experiences. Eudaimonia, its contrast and companion, refers to an orientation towards self-expression, personal growth, and meaningful experiences. Early work focused on hedonia or hedonic experiences with technology \cite{diefenbach_hedonic_2014, hancock_hedonomics_2005, hassenzahl_effect_2001, helander_hedonomicsaffective_2002}. Eudaimonia has only recently begun to experience an uptick in scholarly attention within HCI \cite{deterding_eudaimonic_2014, mekler_momentary_2016, muller_facets_2015, seaborn_evaluating_2016, wirth_beyond_2012,oliver_entertainment_2011}, media \cite{tamborini_defining_2010, bartsch_appreciation_2017}, entertainment \cite{bartsch_appreciation_2017, oliver_entertainment_2011, wirth_beyond_2012}, and game studies \cite{cole_thinking_2021, daneels_eudaimonic_2021,seaborn_eudaimonia_2020}. Yet, eudaimonia has been presumed “positive,” which, like UX, may not reflect experiential realities and desires. In response, philosopher of eudaimonia Waterman coined a new concept in 2021: \emph{maldaimonia} \cite{waterman_toward_2021}. This refers to an orientation towards pleasure and self-expression that is egocentric, exploitative, and/or destructive, or even harmful. People may have maldaimonic experiences with technology, but this is difficult to study because of social taboos \cite{waterman_toward_2021} grounded in well-known biases, including social acceptability biases \cite{krumpal_determinants_2013} in the self and towards others \cite{millham_need_1980}.

One type of interactive technology offers a way forward: video games. Games provide a “magic circle” \cite{salen_rules_2004} that people voluntarily enter wherein the normal rules and norms of reality are suspended. Since many do not see games as “real,” they may be more willing to talk about the “bad” things they have done or experienced \cite{cook_under_2018,kowert_toxicity_2022, mattinen_online_2018}. Yet, gameplay, like UX, is complex and emergent \cite{gundry_validity_2019, gundry_intrinsic_2018}, dependent on the player and the instance of play, as well as the game and other actors therein. Some players engage in \emph{transgressive play} \cite{jorgensen_transgression_2019, mortensen_paradox_2020}, purposefully violating the “magic circle” in a controversial way, such that the play is serious, distressing, or harmful. We do not advocate for such forms of “play,” although it may be an expression of maldaimonia. Some games, like \emph{Doom} (1993), \emph{Thief} (1998), and \emph{Deus Ex} (2000), allow the player to decide how to win: through harmful acts or alternatives. In multiplayer contexts, some partake in \emph{dark play} \cite{mortensen_dark_2015} or \emph{cruel play} \cite{sutton-smith_ambiguity_2001}, a form of deception whereby the player plays by other rules and is deviant. \emph{Dark participation} \cite{kowert_dark_2020,kowert_toxicity_2022} can also occur in social contexts, characterized by toxicity, trolling, bullying, deviancy, and harassment by anonymous actors who can shirk responsibility and hide behind anonymity \cite{kowert_dark_2020,tang_mens_2016}. But “dark” orientations and experiences may not always be \emph{social}, i.e., dark participation and dark play, deceptive, i.e., dark play, or violate the \emph{norms} of play and break the fourth wall, e.g., transgressive play. And they may also be \emph{meaningful} or \emph{self-actualizing}, i.e., eudaimonic. As yet, a clear experiential construct on such player phenomena has not been distinguished.

To this end, we conducted an online critical incident survey on people’s experiences of maldaimonia in games. We followed the first steps taken by Müller, Mekler, and Opwis \cite{muller_facets_2015} and alter Mekler and Hornbæk \cite{mekler_momentary_2016} when initially exploring the experiential nature of eudaimonic UX. Our goal was to take the first step towards finding empirical support for maldaimonia using a tested methodology on a similar construct. We asked: \emph{Is maldaimonic game UX a valid experiential construct?} We discovered that people do have maldaimonic experiences in games and are willing to report on them in detail. Our main contributions are initial empirical validation of maldaimonic UX within games and a set of implications for theorizing and studying player experience from a maldaimonic perspective.

\section{Theoretical Background}

Waterman anchored the concept of maldaimonia on eudaimonia. With Hellenic roots in Aristotle’s \emph{Nichomachean Ethics} (4th century BCE) \cite{waterman_toward_2021}, eudaimonia has been defined as an orientation towards and/or state of happiness (in posterity) and flourishing (in modernity) \cite{deci_hedonia_2008,huta_pursuing_2013,huta_eudaimonia_2013,waterman_personal_1990,waterman_reconsidering_2008,waterman_two_1993}. As an \emph{experiential construct} representing the pursuit of meaningful engagement, eudaimonia is often contrasted with hedonia, or the pursuit of pleasure \cite{huta_pursuing_2010,huta_eudaimonia_2013}. Maldaimonia emerged from a critical lens on the ethics of assigning \emph{negative} forms of orientations, actions, and experiences as “eudaimonic.” Waterman argues that egocentric, destructive, exploitative, and otherwise harmful forms of engagement cannot be called virtuous, good, or desirable, at least in theory. Quoting Haybron \cite{haybron_philosophical_2016}, he questions: “Can the wicked flourish, at least in principle?” If eudaimonia is about \emph{virtue}, could maldaimonia be about \emph{vice}? Waterman outlines four properties of maldaimonia: (1) attaching positive valence to egocentric, destructive, and/or exploitative activities; (2) providing a basis of personal identity; (3) striving for excellence or mastery in these activities; and (4) aligning these activities as acts of personal expressiveness. Thus, in theory, maldaimonia is distinguished from eudaimonia by its ethical properties and by “positive” self-fulfillment and meaningful engagement.

The challenge is how to validate the construct in reality. Asking most people directly about their own maldaimonic behaviour and experiences is difficult. Waterman proposed \emph{video games} as viable for exploration. As he points out, games can and often do provide positive experiences that could be egocentric, destructive, and/or exploitative. Games also require skill-building and identity construction. Games often allow people to express themselves in ways that they may not otherwise be able to do in “real” life. Indeed, hedonia and eudaimonia have been identified as features of game UX \cite{cole_thinking_2021,daneels_eudaimonic_2021,koek_meaningful_2022,kumpel_effects_2017,seaborn_eudaimonia_2020,wang_exploring_2021}. Nevertheless, Waterman argues that maldaimonia in games may be \emph{symbolic} rather than “real.” Games may merely provide a vessel through which one can carry out activities one knows are unacceptable elsewhere, and without repercussions in the “real” world \cite{cook_under_2018}. Yet, we can also argue that this may be an act of Nietzschean \emph{sublimation} \cite{gemes_freud_2009}, or integration of \emph{part of} one’s true self-expression within a viable, perhaps “half-real” arena \cite{juul_half-real_2011}. We are multi-faceted, and our needs, desires, and behaviours are contextual. This may be true for maldaimonia, as well.

Maldaimonic game UX is not unprecedented. Violence in games is a long-standing and unerringly hot topic \cite{anderson_update_2004, ferguson_good_2007}. Cheating is another potentially maldaimonic example with a long history in game studies \cite{consalvo_cheating_2009,passmore_cheating_2020}. Destructive acts, from property damage to killing off NPCs \cite{pimentel_your_2020}, have been far less explored, even though they feature prominently in games; consider \emph{The Sims} (1998), \emph{Catlateral Damage} (2014), or any game that has breaking and displacing built-in. Perhaps even less explored is the effects of taking on identities through what we might call “maldaimonic” characters \cite{grizzard_being_2014}. The body of work suggests that maldaimonic experiences may exist across a variety of games, within and outside of multiplayer contexts, and within the magic circle of play, yet have implications for emotional, social, and ethical engagement.

\section{Methods}
We conducted an online survey using the critical incident method \cite{woolsey_critical_1986}, a qualitative approach to gathering \emph{critical}, i.e., significant and influential, self-reports of \emph{incidents}, i.e., occurrences and experiences, through an interview or questionnaire format. Having a long history in industrial psychology and usability studies \cite{woolsey_critical_1986}, this method has also been used to capture positive and negative experiences with technologies in a post-hoc fashion \cite{hassenzahl_experience_2010,mekler_momentary_2016,muller_facets_2015}. This study was approved by the university ethics committee on August 9\textsuperscript{th}, 2022 (\#2022153). Responses were collected from August 24\textsuperscript{th}, 2022 until August 31\textsuperscript{st}, 2022. Our protocol was registered in advance of data collection on OSF\footnote{\url{https://osf.io/7cvs9?view_only=28e068f2d70b4998a8bc9d8043a5efbd}} on July 5\textsuperscript{th}, 2022.

\subsection{Participants}
We gathered anonymous responses from 51 adult (aged 21+) US participants through Amazon Mechanical Turk (AMT). This sample size is appropriate for the critical incident method (e.g., N=45 \cite{partala_understanding_2012}). We first conducted a pilot study with n=10 participants. Then, 51 participants completed the full study. For quality assurance, we required that all participants had Masters status, granted by Amazon after high performance across a range of tasks\footnote{\url{https://www.mturk.com/worker/help}}, at least a 95\% HIT approval rate, and had purchased a video game (as a baseline measure of game experience). We also used Nicoletti’s verification code procedure\footnote{\url{http://nicholasnicoletti.com/survey-monkey-and-mechanical-turk-the-verification-code}}, which all participants completed correctly. With these stringent criteria, we were able to accept all responses. As such, 24 women (47.1\%) and 27 men (52.9\%) participated (no one of other gender identities). Most identified as white (41, 80.4\%), with two Black and/or African American (3.9\%), two Hispanic and/or Latin American (3.9\%), one Southeast Asian (2\%), and one East Asian (2\%). All participants had a high school education or higher, and most had a degree (29, 56.9\%). Participants were paid in line with the ethics board standards for participant compensation. Since the survey was estimated to take about 20 minutes, we offered USD \$3.70 as compensation via AMT.

\subsection{Procedure}
We provided a link to a Google Form on AMT. At the top of the form, the idea of maldaimonia was introduced: “‘Maldaimonia’ refers to feelings of enjoyment and self-fulfillment through egocentric, destructive, and/or exploitative acts. In games, this could include, for example, killing enemies, stealing from others, and destroying cities. These are just examples; there may be more.” After submitting their responses, they were provided with a participation code for AMT to receive credit for their work. The average completion time according to AMT was 16 minutes and 27 seconds.

\subsection{Qualitative Instrument}
We elicited accounts using open-ended questions \cite{mekler_momentary_2016}. We began by defining maldaimonia, providing some examples, and requesting “one example from your personal experience.” We prompted: "Bring to mind a single ‘maldaimonic’ experience you've had in a game. Think of ‘maldaimonia’ in whatever way that makes sense to you. You can choose any type of game on any platform, including games played on a smartphone, a computer, a console, etc. You can share a solo or multiplayer experience. It can be one where you chose to act in a maldaimonic way or someone else did, or it was required by the gameplay." We asked for the name of the game, the platform it was played on, e.g., smartphone, Nintendo Switch, what motivated them to play the game in general and maldaimonically, to account for the influence of prior motivations \cite{huta_pursuing_2013}, and who they played the game with, if anyone. We then asked respondents to “describe the ‘maldaimonic’ experience you had in the game. Focus on how your experience involved feelings of enjoyment and self-fulfillment through egocentric, destructive, and/or exploitative acts in the game. Please be as detailed as possible.”

\subsection{Data Analysis}
We used hybrid thematic analysis \cite{proudfoot_inductivedeductive_2023} with a combination of deductive and inductive theme development. For the deductive thematic framework, we used the theoretical models underlying the PXI \cite{abeele_development_2020}: Means-Ends theory \cite{gutman_means-end_1982} and the Mechanics, Dynamics, and Aesthetics (MDA) model \cite{hunicke_mda_2004}. This choice was based on a view towards a future quantitative instrument for maldaimonia. The PXI is a reliable and validated measure based on solid theoretical models and conceptually compatible with maldaimonia as an experience. For the analysis, we used the factors making up the psycho-social consequences layer as themes: mastery, curiosity, immersion, autonomy, meaning. We also used inductive theme development as recommended for the critical incident method \cite{woolsey_critical_1986}. For this, the first author considered how the theory of maldaimonia, as outlined, mapped and did not map onto the gathered accounts of experiences. They generated codes by reading through the data several times and then constructing themes, as per Braun and Clarke \cite{braun_using_2006}, around clusters of codes that could not be classified with the deductive themes. These were then classified under higher-order “types.” The second author was brought in to discuss and achieve consensus on the themes; there were no disagreements. No inter-rater reliability test was conducted. The second author was also responsible for categorizing the gaming context data.

\section{Findings}
\subsection{Where and Who in Maldaimonic UX}
\subsubsection{Gaming Context}
Respondents reported 40 unique games across a range of genres. These included: action-adventure (11, 23\%; e.g., the \emph{Grand Theft Auto series}, the \emph{Saints Row} series, \emph{Pokémon Go}), MMO (7, 15\%; e.g., the \emph{World of Warcraft} series, \emph{Eve Online}, \emph{City of Heroes}, \emph{Ultima Online}), RPG (7, 15\%; e.g., the \emph{Elder Scrolls} series, \emph{Elden Ring}, \emph{The Legend of Zelda: Breath of the Wild}), simulation (5, 11\%; e.g., \emph{SimCity}, \emph{Minecraft}), strategy (3, 6\%; e.g., \emph{Civilization}, \emph{Age of Empires II}), and adventure (2, 4\%; e.g., \emph{Moirai}, \emph{Days Gone}), with others in action (e.g., \emph{Uncharted 4}), survival (e.g., \emph{survivor.io}), and stealth (e.g., \emph{Thief}). Others included shooter games and especially first-person shooter or FPS (6, 13\%; e.g., \emph{Doom Eternal}, \emph{Half Life}, \emph{Hotline Miami 2}, \emph{Back 4 Blood}) and battle games, including battle royale and battle arena-style games (3, 6\%; e.g., \emph{Fortnite}, \emph{League of Legends}, \emph{World of Warships}). Platforms included PCs (24, 48\%), PlayStation (15, 30\%) smartphones (5, 10\%), Xbox (4, 8\%), and the Nintendo Switch (2, 4\%). This indicates that maldaimonic experiences can happen across a variety of games and genres and systems, even those not typically associated with violence or “dark play.”

\subsubsection{Social Context}
Most people reported having maldaimonic experiences alone, in single-player play sessions (29, 57\%). This distinguishes maldaimonia from dark participation \cite{kowert_dark_2020,kowert_toxicity_2022} as being more about solo play and individual experience, congruent with Waterman’s criteria for maldaimonia as a construct of identity forming and expression \cite{waterman_toward_2021}. Still, seven accounts related to two-player experiences (14\%), twelve were multiplayer (24\%), and three involved massive numbers of people, hundreds or more (6\%). Of the social experiences, thirteen involved friends or family (62\%), five involved strangers (24\%), and three involved a mix of friends, family, and strangers (14\%). While we might expect people to carry out maldaimonic acts against people that they do not know, i.e., to avoid identification and accountability \cite{nitschinsk_disinhibiting_2022}, or encounter maldaimonic acts with strangers, this was not the case in most accounts. This suggests an element of personal expressiveness deemed socially acceptable, the fourth of Waterman’s criteria situated against common perceptions of play as “not real.”

\begin{table*}[]
\caption{Maldaimonic game UX factors classified into higher order types and linked to the four criteria of maldaimonia.}
\label{tab:themes}
\begin{tabular}{p{52pt}p{70pt}p{315pt}p{30pt}}
\toprule
Type           & Factor             & Description                                                                                                                                  & Criteria    \\
\midrule
Transgressions & Murder \& Mayhem   & Destructive acts, including killing, maiming, and harm, as well as destroying property and/or environments.                                & (1)         \\
               & Chaos              & Actions that lead to confusion, instability, disorder, or simply   have no point or impetus.                                                 & (1)         \\
Reflections    & Rule Subversion    & Getting away with morally and/or ethically deviant actions, including   exploitation, stealing, looting, cheating, insults, and shady deals. & (1),(3)     \\
               & Hubris             & Recognition of pride and confidence, especially extreme pride and   overconfidence.                                                          & (2),(3),(4) \\
               & Vengeance          & Reacting to perceived or actual slights with vengeful acts, often   of escalating severity.                                                  & (1),(4)     \\
Feelings       & Malight            & Malicious delight (“malight”) directly linked to the maldaimonic   experience.                                                               & (1)         \\
               & Power              & Expressions of power and invincibility, and the successful use of   force.                                                                   & (3)         \\
               & Mood Shifts        & Moods and affective states influence or are influenced by   maldaimonia.                                                                     & (2)         \\
Appreciations  & Extrinsic Appetite & Satisfaction of needs and desires through external factors, such   as rewards, praise, fame, and collecting material goods.                  & (3),(4)     \\
               & Aesthetics         & Recognizing or incorporating the visuals, sounds, animations, and   other sensory features.                                                  & (4)        \\
\bottomrule
\end{tabular}
\end{table*}

\begin{table*}[]
\caption{Psychosocial and maldaimonic game UX factors in the critical incident accounts.}
\label{tab:factors}
\begin{tabular}{p{52pt}p{70pt}p{315pt}p{30pt}}
\toprule
Type                 & Factor             & Example                                                                                                                                                                                                                                                                                                                                       & Freq.     \\
\midrule
Psychosocial Factors & Mastery            & P26: “Beating   down their Pokemon methodically and strategically knowing when to use my   shields is the best part.”                                                                                                                                                                                                                         & 30 (59\%) \\
                     & Curiosity          & P41: “Trying to survive for as long as   possible with a 5 star wanted level. I was in a hospital fighting off so many   cops for like an hour. It was so crazy how many cops I beat and how long I   survived for.”                                                                                                                          & 2 (4\%)   \\
                     & Immersion          & P34: “its all   I wanted to do in the game (so much that I forgot to do other parts of the   game that were needed as well)”                                                                                                                                                                                                                  & 4 (8\%)   \\
                     & Autonomy           & P51: “I made   my character steal a car from someone … drove it around the city and crashed {[}it{]}.”                                                                                                                                                                                                                                        & 23 (45\%) \\
                     & Meaning            & P19: “They appeal to my creative side … it's   fun to do something illegal and get away with it.”                                                                                                                                                                                                                                             & 1 (2\%)   \\
\midrule
\multicolumn{4}{l}{Maldaimonic Game UX Factors}\\
\midrule
Transgressions       & Murder \& Mayhem   & P27: “this underhanded and grief-y style of playing where I just murdered curious folks from the safety of my own home.”\newline P37: “As the goat, I would headbutt people into traffic so that they would get hit by cars.” & 40 (78\%) \\
                     & Chaos              & P8: “I just liked going to these areas and killing everyone,   trying to start a faction war/fight.”  & 9 (18\%)  \\
Reflections          & Rule Subversion    & P3: “you could do anything you wanted in the game including running from the police”                                                                                                                                                                                                                                                        & 20 (39\%) \\
                     & Hubris             & P45: “One of my favorite melee weapons is the Scythe (…) It's   considered somewhat insulting to be killed by melee weapons as it implies the   opponent can't aim well to shoot the attacker”                                                                                                                                                & 5 (10\%)  \\
                     & Vengeance          & P22: “I think I went overboard. I just wanted to prove a point   and to completely annihilate the other.”                                                                                                                                                                                                                                     & 3 (6\%)   \\
Feelings             & Malight            & P20: “I derived great satisfaction and enjoyment from mowing   down these rampaging zombies. It felt thrilling to come through large battles   with low odds of success.”                                                                                                                                                                     & 32 (63\%) \\
                     & Power              & P30: “You are the boss of the gang, so everyone looks to this   character for leadership.”                                                                                                                                                                                                                                                    & 27 (53\%) \\
                     & Mood Shifts        & P21: “When I first started the game, I was actually a little   taken aback by some of the violence. … After a while, I started to really   enjoy it. There was almost a sadistic pleasure in being good enough to   destroy an entire village and kill everybody in it. I started to enjoy my   success and feel a little bit vindictive ...” & 12 (24\%) \\
Appreciations        & Extrinsic Appetite & P25: “It makes me feel like I'm going something good when I   kill them because they reward me with loot.”                                                                                                                                                                                                                                    & 17 (33\%) \\
                     & Aesthetics         & P23: “jumping off a high cliff thing and landing a deadly blow   which looked cool.”                                                                                                                                                                                                                                                          & 6 (12\%) \\
\bottomrule
\end{tabular}
\end{table*}

\subsection{Why Play: Patterns of Maldaimonic Orientations and Experiences}
\subsubsection{Motivations to Play}
Four respondents (8\%) indicated that they had played the game with the goal of having a maldaimonic experience. For example, P3 stated, “\emph{freedom to do what you want including illegal acts},” while P10 was inspired by a friend’s account of the maldaimonic experiences that the game offered: “\emph{my friend at work, they would go play it and run into things, people, steal cars and things like that}.” The rest provided a variety of reasons ranging from a desire for fun, stress relief, social experiences, a challenge, and being influenced by others to play. P29, for instance, wrote “\emph{My nephew started me playing it when we were all at the beach on year on a family trip and now we have all been playing this RPG game for years like 8 plus years}.” This suggests that people did not plan or expect to carry out or participate in maldaimonic activities. 

\subsubsection{Motivations to Play Maldaimonically}
Respondents generally indicated that the game made them play in a maldaimonic way (18, 35\%). Otherwise, most sought mastery (12, 24\%), amusement (7, 14\%), were curious about maldaimonic acts and consequences (7, 14\%), driven by extrinsic needs satisfaction (7, 14\%), or were feeling competitive (7, 14\%). On this point, some sought revenge, like P44, who “\emph{wanted to give them a taste of their own medicine},” while others were in it for the warmongering, like P14: “\emph{I am petty for revenge}.” Some were seeking an extraordinary experience (4, 8\%) “\emph{that can’t be done in real life}” (P37). Others were bored (3, 6\%) or looking for an escape from frustration and stress in real life (5, 10\%). A few considered games a moral “\emph{hall pass}” where they could explore maldaimonic situations without repercussions. This speaks directly to the notion of the “magic circle” of games \cite{salen_rules_2004}, a demarcation from “real life” that players voluntarily step into for creative exploration and other pursuits. Notably, no one sought an immersive experience or meaning, the latter of which directly contradicts the expected pattern of maldaimonia as a corollary of eudaimonia but speaks to hedonic and extrinsic desires.

\subsection{What It Is: Maldaimonic and Psychosocial Factors of the Player Experience}
People provided \emph{critical incident accounts} of maldaimonia on a wide array of \emph{game experiences}. We found that these were characterized by the range of psychosocial factors in the PXI \cite{abeele_development_2020} (Table~\ref{tab:factors}). However, we also derived a new set of ten maldaimonic-specific and game-contextualized factors, which we classified under four high-level types as well as linked to the four criteria of maldaimonia as proposed by Waterman (Table~\ref{tab:themes} and Table~\ref{tab:factors}).

The accounts mapped onto the three activities central to Waterman’s definition of maldaimonia. We identified \emph{egocentrism} (Hubris, Vengeance, Mood Shifts), \emph{destruction} (Murder and Mayhem, Chaos, Power), and \emph{exploitation} (Rule Subversion, Extrinsic Appetite). However, two may be particularly relevant to game UX: Malight, or Malicious Delight, and Aesthetics. Games are typically created for entertainment and pleasure, linking the maldaimonic factor of Malight to hedonia. Aesthetics refers to the interactive medium, the modalities through which the game is played, and how these are used to craft visual, audio, and tactile experiences. For example, P9 describes a vivid scene from a \emph{Grand Theft Auto} game: “\emph{you can cause a massive accident that just looks impressive and cars are completely wrecked}.” Additionally, accounts ranged in severity. For instance, P10 recounted feelings of maldaimonia about mischief with paper: “\emph{Honestly, I just found that I really enjoyed running into the news paper stands. It made me happy for some reason}.”

The ethical quandary of maldaimonic actions appeared to be grounded in self-identity and self-expression, two of Waterman’s criteria. Some expressed a laissez-faire attitude, such as P6, who “\emph{I got dressed and got in my car with my guns and just went around being a menace}.” Others offered a defense, an example self/other reputation management \cite{millham_need_1980}. P37 explained that the simulated environment was a caveat: “\emph{Being so transgressive in such a ridiculous way was hilarious and kind of thrilling, but only because it wasn't real people dying}.” P14 ruled that the other player was a “griefer,” someone who deliberately provokes others to ruin their enjoyment of the game, and so they were ethically exempt: “\emph{It was clear that he was a griefer as I saw other people talking about him in the lobby}.”

Only a few accounts (8, 16\%) related to dark participation, involving negative human-to-human interactions. The relative scarcity of bullying, toxicity, and abuse in our corpus suggests that maldaimonic is a distinct concept from dark participation, even while by some accounts it overlaps with features of “toxic game spaces.”

\section{Discussion}
Maldaimonic experiences happen in games. The array of accounts indicate that maldaimonic game UX is a legitimate experiential construct grounded in player psychosocial needs and consequences. Moreover, it relates in ways both expected and unexpected with its theorized “good” twin, eudaimonia, as well as features of the game experience, including player experience, moods and affective states, and mechanics. Maldaimonic game UX, as indicated by the findings on our corpus of critical incident self-reports, is a distinct concept. Overlapping to a small degree with dark participation \cite{kowert_dark_2020,kowert_toxicity_2022} and transgressive play \cite{jorgensen_transgression_2019,mortensen_dark_2015}, it appears to be centred on solo player experiences that allow for immersive self-expression without an audience or where the “transgression” is socially acceptable by known peers and social networks. Critical incidents were reported across a variety of game genres and particular games, suggesting that it is not tied to specific game experiences, notably, violent games, despite what Waterman’s contemplations and the “dark play” literature might predict. Nevertheless, maldaimonic experiences were overwhelmingly positive ones. People tended not to feel angry, guilty, or sad. People played maldaimonically when required by the goals and requirements of the game itself, but a large subset also sought mastery, pleasure, and the fulfillment of extrinsic needs through egocentric, destructive, and/or harmful acts. Indeed, this suggests that we should explore the roles of hedonia, i.e., pleasure \cite{huta_pursuing_2010}, and extrinsic motivation and needs satisfaction \cite{ryan_intrinsic_2000} within maldaimonic (as well as eudaimonic) game experiences.

All four of Waterman’s \cite{waterman_toward_2021} proposed criteria for maldaimonia as an experiential construct were found in the critical accounts of gaming experiences. As Table~\ref{tab:themes} demonstrates, there are varieties of maldaimonic game UX, such that diverse experiences of play may not necessarily have the same maldaimonic features, even if they meet the four criteria for maldaimonic play. For example, experiences characterized as Malight and Hubris (two of the ten factors across two higher order types) would suffice. We can generalize the findings with respect to Waterman’s criteria. The first criteria, attaching \emph{positive} valence, emphasis on the “positive,” to egocentric, destructive, and/or exploitative activities, was largely confirmed. Indeed, some experienced a form of affective synchrony \cite{rafaeli_affective_2007} through engagement in the maldaimonic experience, feeling better by playing a game that matched their mood at the time. The second, providing a basis of personal identity, was more nuanced, with many reports of pride, perhaps hubris, and self-attribution of mastery and accomplishments alongside feelings of incongruity. The third, striving for excellence or mastery, was the most frequent psychosocial factor of experience, and thus confirmed. The last, assigning these activities as acts of personal expressiveness, depended on the person. Autonomy was the second most frequent psychosocial factor, but the perceived benefits of the expressiveness that this feeling of autonomy offered varied. Perhaps maldaimonia is not universal, but it does exist.

We can summarize the features of maldaimonic game UX as:

\begin{itemize}
    \item Found in single-player or, collaborative multiplayer, and/or social contexts with known others
    \item Confined to the “magic circle” of the game
    \item Existing across a variety of games and game genres
    \item Marked by positive affect and psychosocial consequences
    \item Engendered through multiple combinations of maldaimonic factors linked to Waterman’s four criteria
\end{itemize}

Going forward, standardized methods of evaluating and/or measuring maldaimonia will be needed. Waterman \cite{waterman_toward_2021} proposed the development of a two-factor instrument, including personal expressiveness and ethicality scales. Schadenfreude, translated from the German as “malicious delight,” has been included in previous game instruments \cite{de_kort_digital_2007}; items could be extracted and expanded upon to develop a comprehensive scale. Here, we contribute a multi-tiered thematic framework of maldaimonic game UX; this framework could also seed the creation of a quantitative instrument.

\subsection{Limitations}
A diversity of gamers, especially non-white people beyond the US, will need to be targeted in future. Post hoc recall is limited \cite{huta_pursuing_2010,kim_pleasure_2014,mekler_momentary_2016}; since we cannot predict when maldaimonic experiences will occur, this is likely to continue being a challenge.

\section{Conclusion}
Maldaimonic UX appears be an experiential construct, at least in the context of games. There is ample opportunity to explore whether and how maldaimonia in games relates (or not) to maldaimonia in other contexts. We have provided an initial set of empirical findings and descriptive frameworks anchored to critical incident reports of maldaimonic experiences in games. These can be carried forward in the design and study of game UX and UX more broadly.

\begin{acks}
This work was funded by department funds. Our gratitude to Peter Pennefather for years of engaging discussion on eudaimonia and more recent lively debates on maldaimonia.
\end{acks}

\bibliographystyle{ACM-Reference-Format}
\balance
\bibliography{main.bbl}










\end{document}